\documentclass{article}
\usepackage[utf8]{inputenc}
\usepackage{graphicx}
\usepackage{amsmath}
\usepackage{multicol}
\usepackage{geometry}
\usepackage{subfig}
\usepackage{gensymb}
\usepackage{caption}
\usepackage{textcomp}
\usepackage{authblk}
\title{Characterisation of a deep-ultraviolet light-emitting diode emission pattern via fluorescence}
\author{Mollie McFarlane* and Gail McConnell}
\affil[*]{mollie.mcfarlane@strath.ac.uk}
\affil{Department of Physics, University of Strathclyde, SUPA, Glasgow, U.K.}
\date{November 2019}

\begin{document}

\maketitle

\begin{abstract}
Recent advances in LED technology have allowed the development of high-brightness deep-UV LEDs with potential applications in water purification, gas sensing and as excitation sources in fluorescence microscopy. The emission pattern of an LED is the angular distribution of emission intensity and can be mathematically modelled or measured using a camera, although a general model is difficult to obtain and most CMOS and CCD cameras have low sensitivity in the deep-UV. We report a fluorescence-based method to determine the emission pattern of a deep-UV LED, achieved by converting 280 nm radiation into visible light via fluorescence such that it can be detected by a standard CMOS camera. We find that the emission pattern of the LED is consistent with the Lambertian trend typically obtained in planar LED packages to an accuracy of 99.6$\%$. We also demonstrate the ability of the technique to distinguish between LED packaging types.
\end{abstract}
\newpage


\section{Introduction}

Recent developments in light-emitting diode (LED) technology have produced deep-ultraviolet aluminium gallium nitride (AlGaN) LEDs with wavelengths ranging between 220-280 nm emitting in the 100 mW range \cite{Kneissl}. These LEDs have applications in sterilisation, water purification \cite{Song} and gas-sensing \cite{Li}. Deep-UV LEDs also have potential applications as excitation sources in fluorescence microscopy. In particular, 280 nm LEDs have an electroluminescence spectrum which overlaps well with the excitation spectrum of many fluorophores including semiconductor quantum dots, aromatic amino acids tryptophan and tyrosine \cite{Lakowicz2} and even standard dyes such as eosin, rhodamine and DAPI \cite{Muse} \cite{MUVE}. 
\\ However, one of the major weaknesses of using deep-UV LEDs for microscopy is low transmission through glass, making deep-UV illumination sources difficult to adapt into standard epifluorescence microscopes. An alternative is to use off-axis critical illumination of the sample \cite{Wong} which can also provide a shadowing effect to enhance tissue surface topography \cite{Muse}. Even more simply, the LED can be used to directly illuminate the specimen without use of any lenses, but this begs the question of homogeneity of specimen illumination and hence the emission pattern of the LED in air is of interest when assessing the suitability of an LED as an excitation source in optical microscopy. \\ The emission pattern of an LED is the emission intensity measured as a function of angle from the normal of the LED chip \cite{Moreno}. In high refractive-index planar LED chips, the refractive index change between LED and surrounding material often leads to a Lambertian emission pattern \cite{Schubert}. The intensity of light in air is described by:

\begin{equation}
 I={\frac{P_{LED}}{4 \pi r^{2}}}
 {\frac{n_{air}^{2}}{n_{LED}^{2}}} cos(\theta)
\end{equation}

 Where $P_{LED}$ is the radiant power of the LED, r is the distance from source to target, $n_{air}$ is the refractive index of air, $n_{LED}$ is the refractive index of the semiconductor and $\theta$ is the angle of emission measured from the normal of the LED chip. Other LED packaging geometries include hemispherical, which exhibit isotropic emission patterns, and parabolic, which have strongly directional emission patterns \cite{Schubert}. Some mathematical models have been developed to simulate the emission patterns of packaged LEDs and LED arrays \cite{Moreno} \cite{Sheng}, however, packaged LEDs come available in different emission patterns which makes a general model difficult to obtain \cite{Moreno} \cite{Taghipour}. Experimentally, the emission pattern can be obtained using a camera, however standard CMOS and CCD cameras are not sensitive in the deep-UV due to glass or poly-silicon elements which absorb in the UV. Back-thinned CCDs offer a solution to this problem, but are costly. Previous work studying AlGaN in LEDs has been shown to overcome these detection limits by using fluorescence to convert deep-UV radiation into visible light \cite{Lapeyrade}. In this work, the authors use fluorescence to image the active region of the chip and investigate the electroluminescence distribution on the microscopic scale. Here, we apply an adaptation of this technique to a 280 nm LED in order to measure the far-field emission pattern of the LED. In this method, the LED is used to directly illuminate a fluorescent specimen which, due to its longer emission wavelength, can be imaged by a standard CMOS or CCD camera. The intensity across the fluorescent specimen is obtained as an indirect measurement of the emission pattern of the LED. This method is applied to measure both the intensity distribution across a 4.5mm field of view, and, by rotating the LED about its axis, the angular emission pattern.

\section{Materials and Methods}

\begin{figure}
\centering
\captionsetup{justification=centering}
 \includegraphics[width=4in]{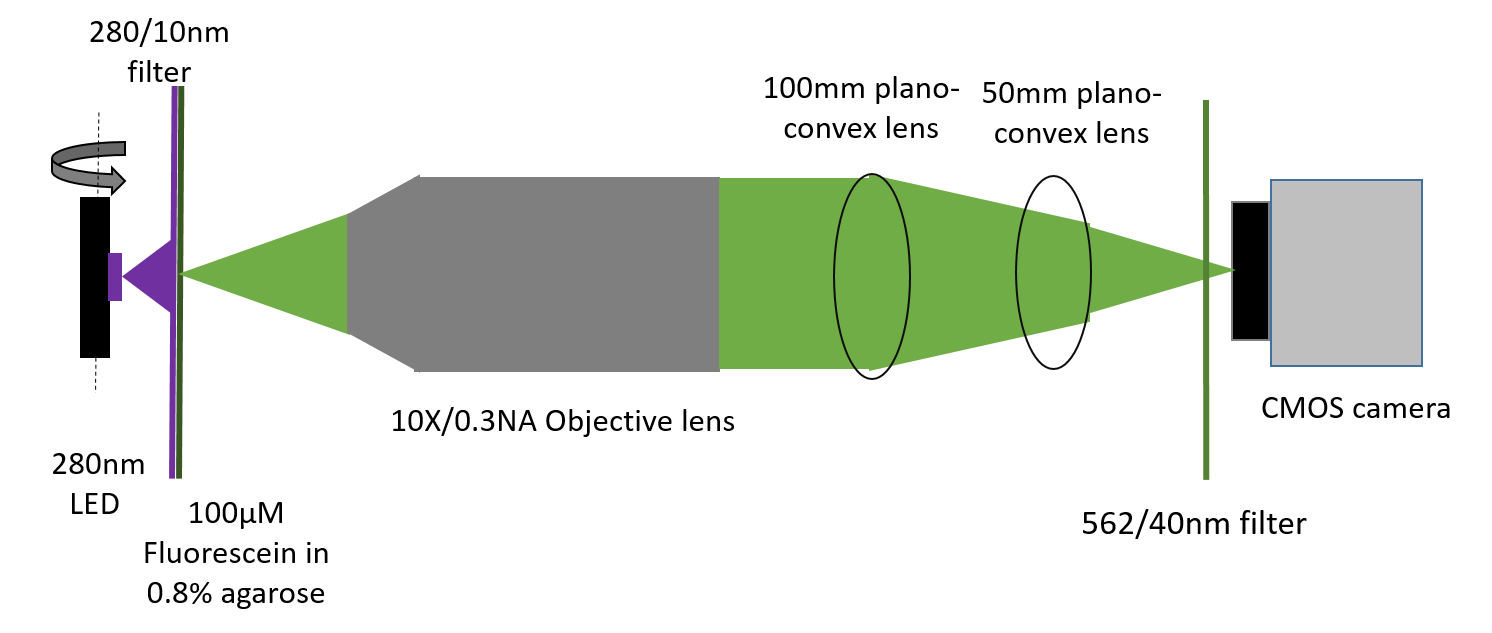}
\caption{\textit{Experimental set up for imaging the emission pattern of a deep-UV LED. A fluorescent sample is placed in the focal plane of the objective lens and the LED is placed adjacent to the sample, behind a bandpass filter which narrows the electroluminesence spectrum of the LED. Two plano-convex lenses are employed to focus the image onto the camera chip. An emission filter is used to exclude any excitation light. To capture angular measurements, the LED was rotated about the indicated axis. }}
    \label{fig:1}
\end{figure}

The experimental set-up to obtain the emission pattern of a 280 nm LED (LG Innotek LEUVA66H70HF00) is shown in figure \ref{fig:1}. The LED was mounted in a platform mount (Thorlabs PCM) and a 280/10 nm bandpass filter (Edmund Optics, 35-881) was placed in front of the LED to narrow the electroluminescence spectrum. 
\\ A fluorescent agarose block was developed using fluorescein due to its broad excitation spectrum extending into the UV and high quantum yield (95$\%$ \cite{Lakowicz}). To develop the agarose block, 0.8\% agarose (Sigma Aldrich 05066) was added to 100 $\mu$M fluorescein (Sigma Aldrich F6377) in distilled water, microwaved until dissolved and pipetted into a 3D printed mould 2 mm in depth attached to a microscope slide. A path length of 2mm was chosen in order to provide sufficient fluorescent signal. The specimen was placed in a slide holder (Thorlabs XF50) as close as possible to the LED. \\
A 10x 0.3 NA objective lens (Olympus UPLFLN10XP) was chosen to collect the fluorescence emission and the focal plane was set to the specimen. Two plano-convex lenses, 100 mm (Thorlabs LA1509-A) and 50 mm (Thorlabs LA1608-A) in focal length respectively, were used to focus the image of the fluorescent sample onto the CMOS camera (IDS UI-3060CP) whilst reducing the beam diameter, resulting in a total magnification of 2.5x and a field of view of 4.5 mm.


An emission filter (562/40 nm, Semrock FF01-562/40-25) was placed in front of the camera. \\ 
Images were acquired at LED drive currents ranging from 50-350 mA. The resulting images were imported into Fiji \cite{Fiji} and background corrected by subtracting an image of the ambient light surrounding the optics. An intensity line profile was taken horizontally across the field of view as this allowed measurement of a larger area due to the aspect ratio of the camera. A linewidth of 50 pixels was taken, corresponding to 116 $\mu$m, to reduce noise in the intensity profile.\\ 
The emission pattern was then measured at different angles to investigate the intensity as a function of angle. To do this, the LED was mounted on a rotating stage (Thorlabs RP01) 65 mm from the sample such that the chip lay on the axis of rotation. The stage was rotated between 0$\degree$ and 90$\degree$ from the normal in both directions at a constant current of 350 mA and images were taken at each angle with a longer exposure time of 50 ms to compensate for the decrease in intensity due to increased LED-sample distance. Image intensity was recorded and plotted as a function of angle. An average of 3 images was recorded per angle. The data was fitted against a perfect Lambertian using equation 1. \\ To test the ability of the technique in distinguishing between types of LED packaging geometries, measurements were repeated with a second deep-UV LED specimen (Thorlabs M275D2) emitting at 275 nm which exhibits a non-Lambertian emission pattern. Images were obtained at a higher exposure time of 150 ms to compensate for the lower radiant power of this LED.

\section{Results and Discussion}

\begin{figure}[h!]
    \centering
    \captionsetup{justification=centering}
    \subfloat(a){{\includegraphics[width=2.5in]{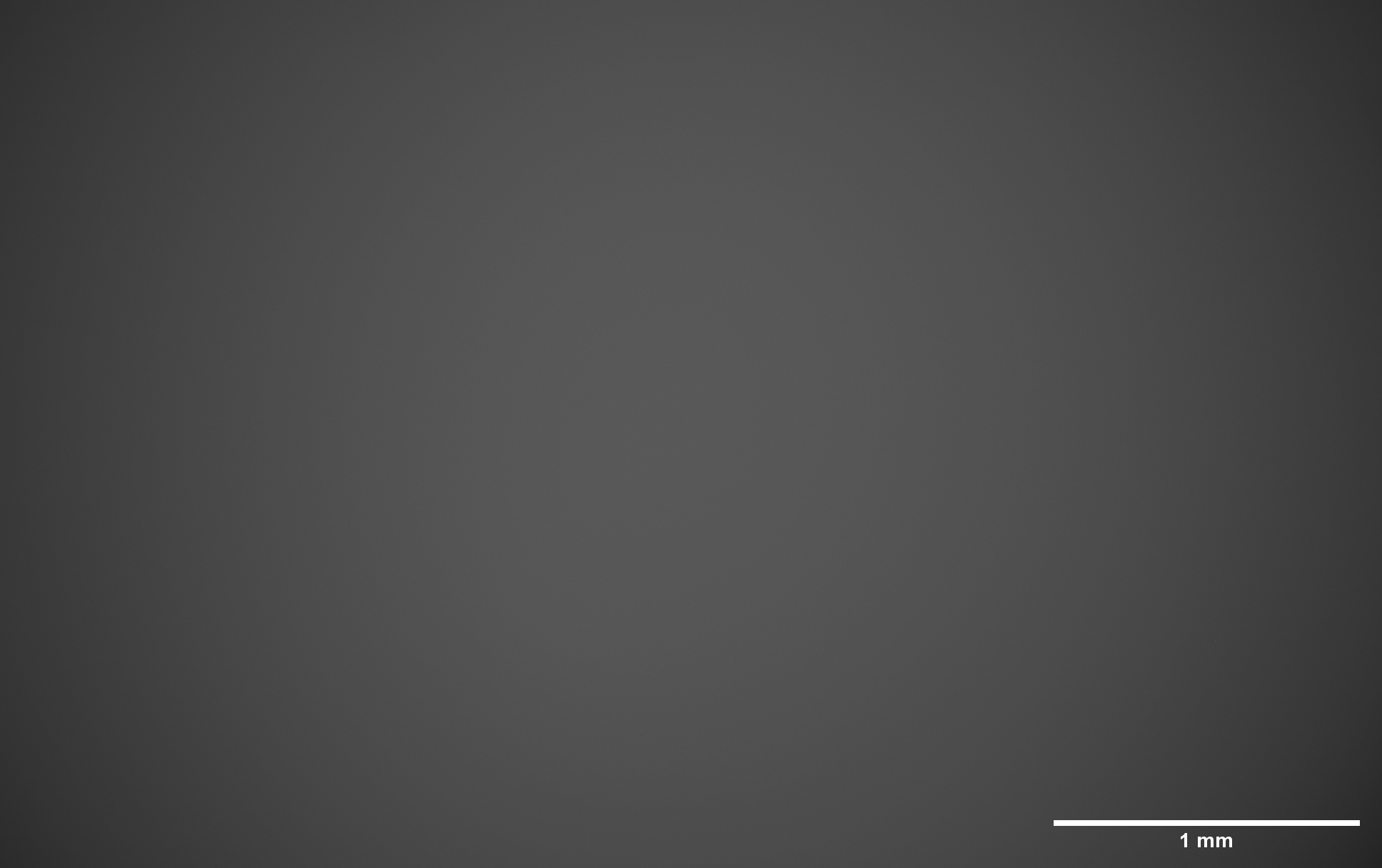} }}
    \qquad
    \subfloat(b){{\includegraphics[width=2.5in,height=1.65in]{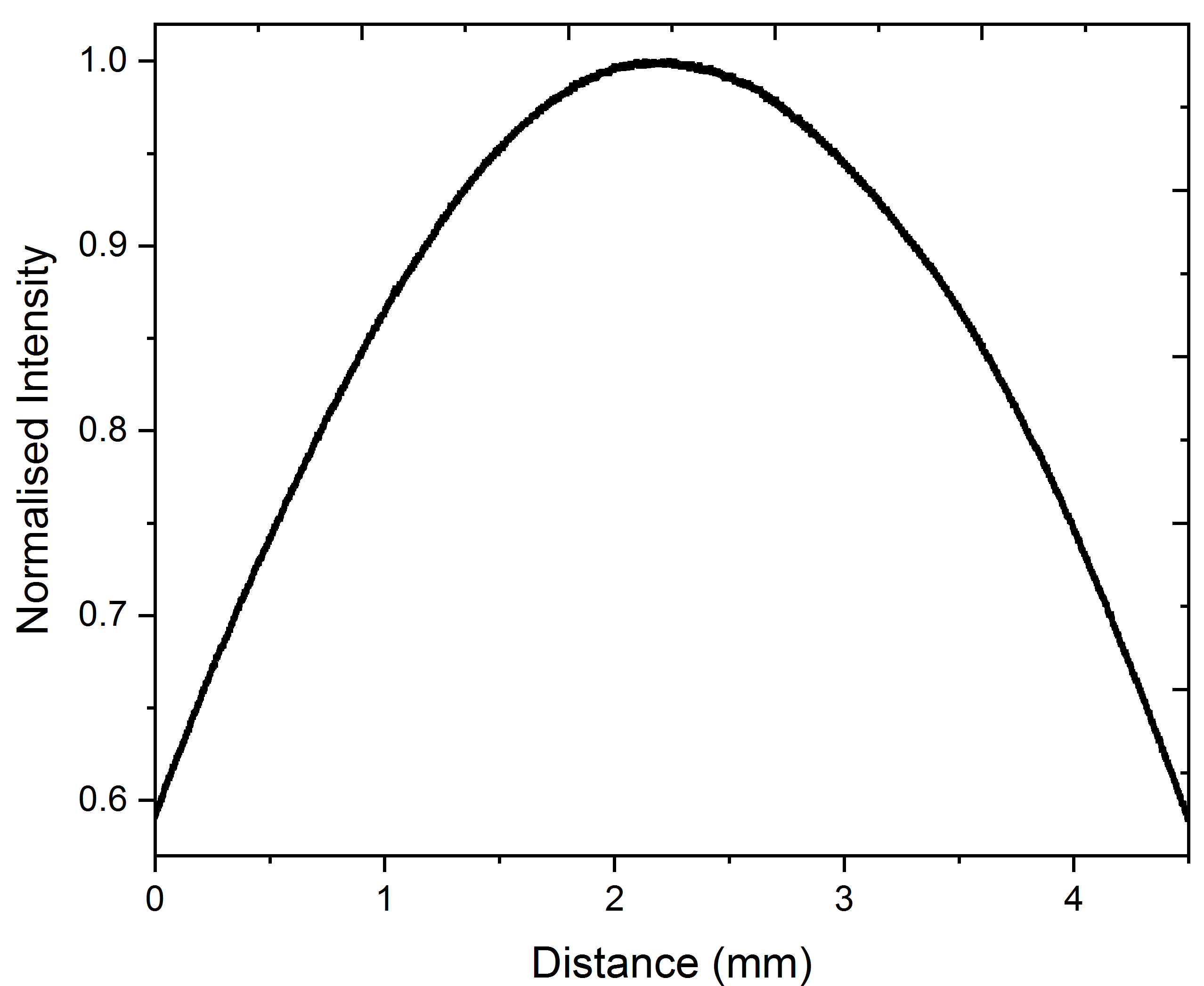} }}%
    \caption{\textit{(a) Image of emission pattern taken at a drive current of 350 mA and camera exposure time of 5 ms. (b) Plot profile of above image taken horizontally across the field of view at a width of 116 $\mu$m}}
    \label{fig:near}%
\end{figure}

The normalised intensity distribution across the field of view of the microscope is shown in figure \ref{fig:near}. 

The fluorescence intensity is at its highest in the center of the field of view, reaching 60\% of its maximum output at the edge. It was found that outside of changes in intensity, adjusting the drive current had no effect on the intensity distribution. Increasing the LED-specimen distance can reduce the variation in intensity as the angular distribution across the field of view becomes smaller. However, this is at the expense of the intensity of light reaching the specimen which is inversely proportional to the square of the LED-specimen distance (equation 1). \\

\begin{figure}[t!]
    \centering
    \captionsetup{justification=centering}
    \includegraphics[width=3.25in]{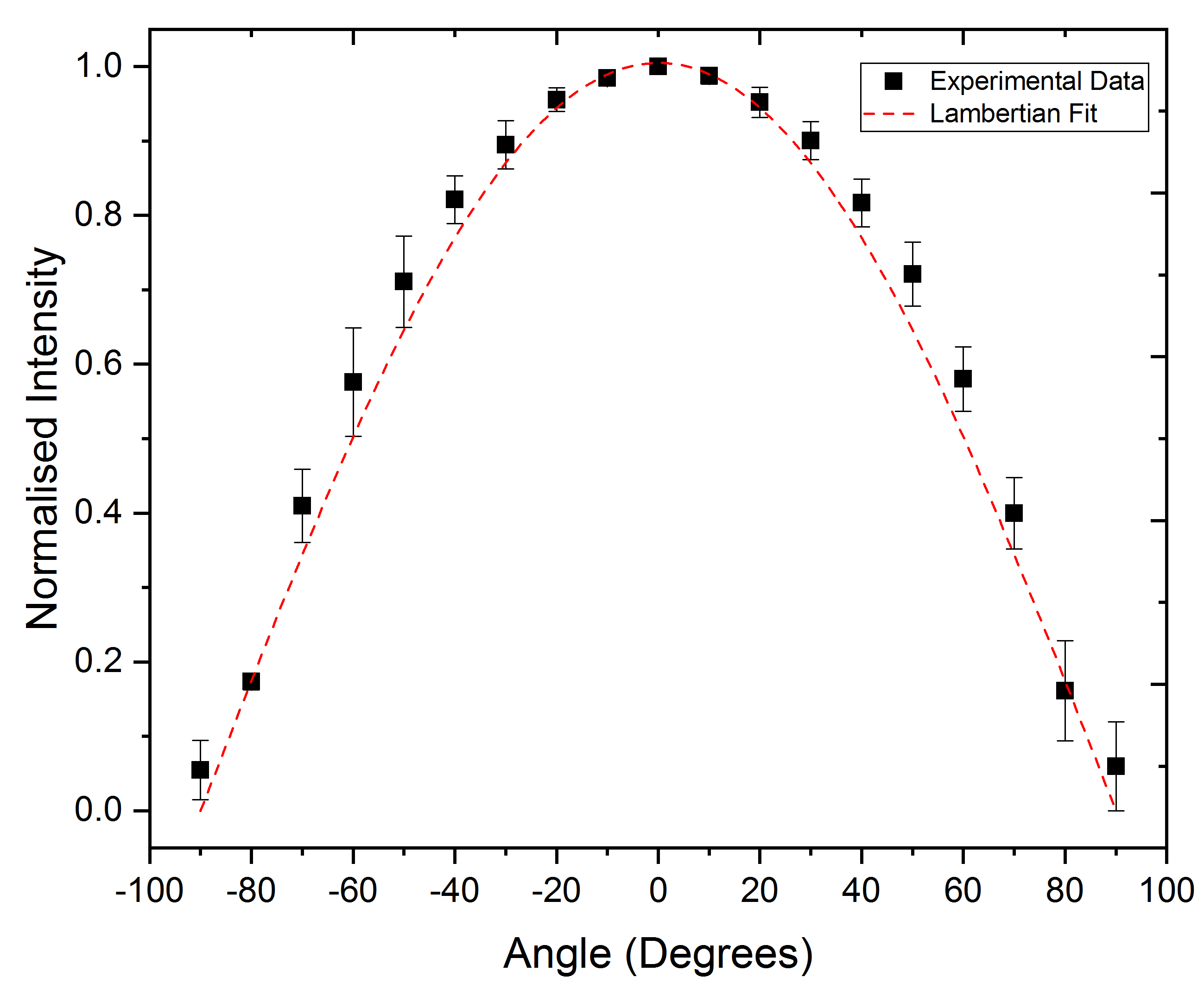}
        \caption {\textit{Emission intensity as a function of angle. The LED was rotated between 0$\degree$ and 90$\degree$ from the normal of the chip surface and each intensity value was recorded. Standard deviation error bars are shown on the y-axis and x-axis error bars correspond to the angular divisions on the rotating stage at $\pm$ 1$\degree$}}
    \label{fig:angle}
\end{figure}

Due to restrictions set by the field of view of the objective lens, this distribution only accounts for a small section of the LED's emission pattern (4.5 mm) which typically extends much further. To account for this, the LED can be rotated around the x-axis to measure the emission pattern at different incident angles. This technique allows for a broader measurement of the emission pattern of the LED and results are shown in figure \ref{fig:angle} along with standard deviation error bars on the y-axis. The incident angle has a tolerance of $\pm$ 1$\%$ as limited by the scale on the rotating stage. The intensity of the fluorescence emission is at its maximum at 0$\degree$ and reaches approximately 50 $\%$ of its maximum value at 60$\degree$ from the normal which is consistent with a Lambertian trend \cite{Schubert}. To test the correlation between data and a Lambertian trend, the data was fitted using equation 1 (shown in dotted lines). The data fits the Lambertian trend with a coefficient of determination of 99.6$\%$, confirming that there is a strong correlation between the experimental data and a Lambertian trend. 

\begin{figure}[t!]
    \centering
    \captionsetup{justification=centering}
    \includegraphics[width=3.25in]{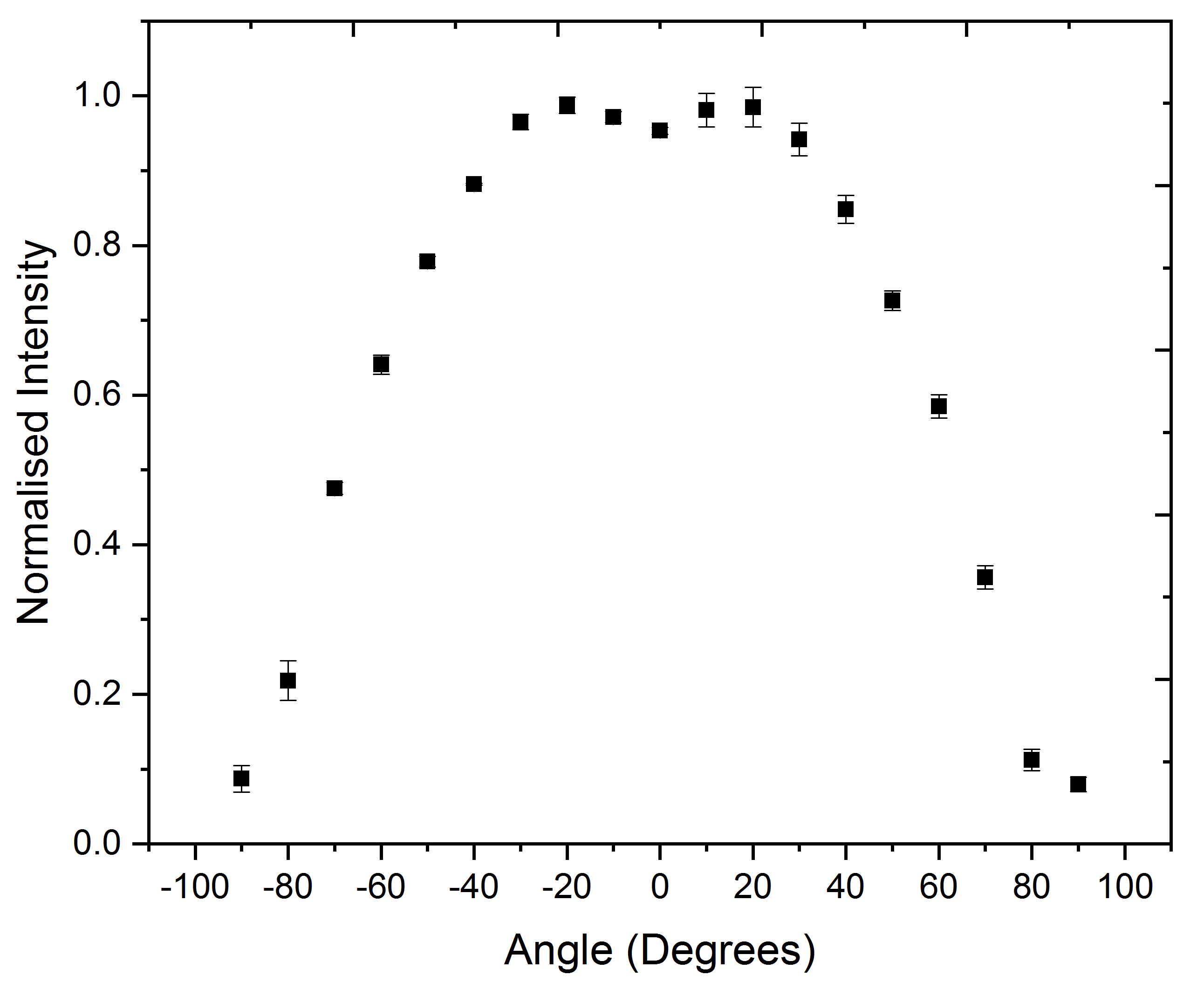}
    \caption {\textit{Emission intensity as a function of angle of a non-Lambertian emitter. Standard deviation error bars are shown on the y-axis and x-axis error bars correspond to the angular divisions available on the rotating stage at $\pm$ 1$\degree$}}
    \label{fig:batwing}
\end{figure}

Finally, a second LED specimen emitting at 275 nm with a non-Lambertian emission pattern was measured to investigate the reliability of the method to distinguish between LED types (figure \ref{fig:batwing}). This LED exhibits an emission pattern towards the batwing shape, with intensity peaking at $\pm$ 20$\degree$. This result is consistent with the data sheet supplied with the LED and confirms that the technique can be used to characterise LEDs with different packaging geometries.

This is an indirect technique which lies on the assumption that fluorophores are distributed evenly across the fluorescent sample and hence changes in fluorescence intensity are directly correlated to changes in LED intensity.  Because of the large field of view desired to measure as large an emission area as possible, the magnification was decreased to 2.5x and images were not sufficiently sampled by the camera according to the Nyquist criteria \cite{Pawley}. \\
As resulting images are convolutions of the LED emission and fluorescence emission, deconvolution is desirable to revert the image to the emission of the LED only. To demonstrate this, the point spread function (PSF) of the microscope can be acquired and blind deconvolution performed to compensate for the increase in full width half maximum (FWHM) caused by the convolution of the excitation light and fluorescent specimen. However, as Nyquist sampling could not be achieved with this set up, a diffraction limited spot would appear as a single bright pixel. To overcome this, the set up was reverted to a higher magnification by replacing the two plano convex lenses with a 200 mm tube lens, resulting in a pixel size of 416 nm which satisfies the Nyquist criteria. Sub resolution fluorescent beads (Fluoresbrite YG Microscopheres 0.5$\mu$ m, Polysciences Inc)  on a quartz microscope slide were used to measure the PSF of the microscope. The image was then subjected to blind deconvolution using a Born and Wolf generated PSF (PSF Generator \cite{PSF}) and a Richardson-Lucy algorithm with 10 steps (DeconvolutionLab2 \cite{Deconvolution}). The FWHM reduced from 1125.3 $\pm$ 52.8 nm to 823.1 $\pm$ 28.6 nm which, in the original set up used for measurement of the emission pattern, would be a change of less than one pixel. \\
Another limitation of the technique is the field of view restriction set by the objective lens which limits the detection range to 4.5 mm whilst the emission pattern of the LED is expected to be much larger. This can be overcome by rotating the LED to measure the full emission pattern, as demonstrated previously using visible spectrum LEDs \cite{Angles}, or scanning the LED in x-y across the field of view. For applications in optical microscopy, 4.5 mm could be considered a sufficient measurement as a typical field of view does not extend far beyond this size. \\
This technique could be applied to measure any wavelength of deep-UV LED due to the ability of fluorescein to become excited by wavelengths as low as 200nm \cite{Gutierrez}. The technique could aid in the design of LED chips and LED arrays where a particular emission pattern is desired, as well as the characterisation of packaged LEDs demonstrated here. In addition, the technique can be used to measure the homogeneity of deep-UV illumination using methods such as critical and K\"ohler illumination by adding the appropriate optics in the illumination pathway between LED and fluorescent specimen, as previously demonstrated using visible wavelength LEDs \cite{Enquist}.

\section{Conclusion}
We have demonstrated a technique capable of measuring the emission pattern of a deep-UV LED on a camera without the requirement of UV-extended sensors. In this method, the LED is used to directly illuminate a fluorescent sample which, due to its longer emission wavelength, can be imaged by a standard CMOS or CCD camera. The intensity across the fluorescent sample is obtained as an indirect measurement of the emission pattern of the LED. The emission pattern was found to follow a Lambertian trend to an accuracy of 99.6$\%$. The technique was successful in distinguishing between emission patterns of Lambertian and non-Lambertian emitters, confirming that the technique can allow for accurate characterisation of LEDs. The technique may also be useful in the design of LED chips or LED arrays where a particular emission pattern is required or to test the homogeneity of critical or K\"ohler illumination methods with deep-UV LEDs for applications in optical microscopy.

\section{Acknowledgements}
This work was supported by Medical Research Scotland (PhD-1157-2017), CoolLED Ltd and the Medical Research Council, grant number MR/K015583/1.

\bibliographystyle{unsrt}
\bibliography{Paper.bib}

\end{document}